\documentstyle[11pt,aaspp4]{article}

\begin{document}

\title{Formation and Evolution of Galactic Halos in Clusters of Galaxies}
\author{Takashi Okamoto \and Asao Habe}
\affil{Hokkaido University, Sapporo 060, Japan}%
\authoremail{okamoto@phys.hokudai.ac.jp}

\begin{abstract}
We investigate effects of time evolution of a rich cluster of galaxies 
on its member galactic halos in the standard cold dark matter (SCDM) 
universe using high resolution N-body simulations. 
We identify several hundred galactic halos within virial radius of our 
simulated cluster. 
We also find that a large number of halos have been tidally disrupted 
at $z = 0$.  
Therefore we improve a method of deriving merging history trees of 
galaxies taking account of tidally stripped galaxies.


The main results are as follows:
(1)  At high redshift ($z \simeq 2$), the mass function of the galactic halos  
which are in the cluster at $z = 0$ is very similar to that obtained in the 
field region and well agrees with the Press-Schechter mass function. 
(2) The mass function of cluster galaxies which consist of both galactic halos 
and tidally stripped galaxies has hardly evolved since $z \simeq 2$. 
This mass function at $z = 0$ is well represented by the Press-Schechter mass 
function at $z = 2$. 
(3) At high redshift ($z > 3$), in the region which becomes the cluster the 
fraction of galaxies which have undergone recent merger is lager than that 
in the field. After $z \sim 3$, however, it rapidly decreases and becomes 
smaller than that in the field.   
(4) The strongly stripped galaxy fraction of the cluster galaxies 
begins to increase from $z \simeq 0.5$. At $z = 0$, a clear correlation appears 
between this fraction and the distance from the center of the cluster.  
(5) Tidally truncated halos have steeper outer profiles than 
those of the model of Navarro, Frenk, \& White (1996, 1997).  
\end{abstract}


\section{Introduction}
It is well established that galaxy populations vary with the 
density of neighbouring galaxies in clusters of galaxies (Dressler 1980) and 
depend on distance from the center of clusters of galaxies 
(Whitmore et al. 1993). 
The increase in the fraction of blue, star-forming cluster galaxies 
with redshift (Butcher \& Oemler 1978, 1984a, b) also has been well 
established. 
Several physical processes have been proposed to explain theses effects, 
including shocks induced by rampressure from the 
intracluster medium (Bothun \& Dressler 1986; Gavazzi \& Jaffe 1987), 
effects of cluster tidal field (Byrd \& Valtonen 1990), 
galaxy-galaxy interactions (Barnes \& Hernquist 1991, Moore et al. 1996, 
Moore, Katz, \& Lake  1998), and mergers of individual galaxies in the 
hierarchical clustering universe (Kauffmann et al. 1993; Kauffmann 1996; 
Baugh et al. 1996). 

The purpose of our study is to investigate  effects of galaxy-galaxy 
and galaxy-cluster interactions on  cluster member galaxies and to 
investigate when these interactions become important during the cluster 
evolution.
As the first step for this purpose, we make cosmological N-body simulations
and study how and when galactic dark halos are 
affected by these interactions. 
In particular, we pay our attention to   evolution of  large galactic 
halos ($M_{\rm h} \ge 10^{11} M{\odot}$). 

Unfortunately,  previous almost dissipationless numerical simulations have been 
failing to follow the evolution of galactic halos in dense environments 
such as galaxy groups and clusters owing to their low-resolution
(White 1976; van Kampen 1995; Summers et al. 1995; Moore, Katz, \& Lake 1996). 

To avoid this apparent erasing of substructures in the dense environments 
known as the "over merging problem", many approaches have been done. 
For example, Couchman \& Carlberg (1992) tagged particles in galactic 
halos before a cluster forms, then applied a halo-finding algorithm only 
for the tagged particles at final epoch. Another idea is to introduce 
"artificial cooling" in a collisionless simulation by collecting 
particles in their collapsing regions into  more massive super-particles 
(van Kampen 1995). 
By these approaches, however, we cannot explore strength of 
 galaxy-galaxy and galaxy-cluster interactions. 
Therefore, we use high resolution N-body simulations and the improved method of 
tracing galaxies to investigate those interactions. 

We should consider hydrodynamic processes of baryonic component, 
because radiative cooling allows baryonic component to sink into the center 
of a dark matter halo where it forms a compact and tightly bound stellar 
system which is hardly destroyed by the tidal force and helps its host halo 
to survive to some degree (Summers et al. 1995). 
However, hydrodynamic simulations, e.g.  smoothed particle 
hydrodynamics (SPH) simulations, need much more 
CPU time than the collisionless simulation. 
Then, it is difficult to perform wide dynamical range simulations by this
approach. 
Therefore, we restrict ourselves to follow  evolution of dark matter halos.
We  trace not only  surviving halos but also  strongly stripped halos 
which may survive as galaxies if   hydrodynamic process  are considered,
because we find many strongly stripped halos in a cluster of galaxies in this paper.

Recently, Ghigna et al. (1998) have reported results of similar simulations  
to ours, independently. 
However, there are two large differences between our study and theirs. 
The first difference is  that the mass of their cluster is about half of ours. 
Therefore our galactic halos suffer influence of 
denser environment and our cluster forms at lower redshift than theirs. 
The second difference is that they investigated the evolution of the cluster 
halos from $z = 0.5$ to $z = 0$. On the other hand, 
we investigate it before the formation epoch of  galactic halos to present time. 
Clearly, our investigation gives  more  information 
about galaxy-galaxy and galaxy-cluster interactions which affect  evolution of galaxies.

In Section 2, we present the method of numerical simulations, our halo-finding 
algorithm and the algorithm to create  halo merging history trees.
The algorithm to create  halo merging history trees is improved 
to handle the galactic halos in very dense environments. 
Our results are presented in Section 3 and are discussed in Section 4. 

\section{Simulation}
\subsection{The simulation dataset}
We first describe two simulations used in this paper, and their specific 
purpose. The overall parameters and mass of the most massive virialized  
objects at $z = 0$  in both simulations are listed in Table \ref{data}.
The back ground model for both simulations is the  standard cold dark matter 
(SCDM) universe with the Hubble constant $H_0 = 100 h$ km/s/Mpc, 
where $h = 0.5$. 
This model is normalized with  $\sigma_8 = 1/b$, where $b = 1.5$. 
%
The simulation A represents to  an average piece of the universe 
corresponding to the ``field" environment within a sphere of radius 7 Mpc 
and we use it to check our halo-finding algorithm and compare with another 
simulation. 
%
In the  simulation B, we adopt the constrained random field method to generate  
the initial density perturbation field in which a rich cluster is formed at 
the center of a simulation sphere of radius 30 Mpc (Hoffman \& Ribak 1991). 
The constraint which we impose is the $3 \sigma$ peak with the 8 Mpc Gaussian 
smoothed density field at the center of the simulation sphere.
To get enough resolution with relatively small number of particles, we 
use the multi-mass initial condition for the simulation B 
(Navarro, Frenk, \& White 1996, Huss et al. 1997). 
This initial condition is made as follows.

First, only long wave length components are used for realization
of initial perturbation in the simulation sphere using $\sim 10^5$ 
particles, and then we perform a simulation with these low resolution 
particles. 
After this procedure, we tag the particles which are inside a sphere of   
radius 3 Mpc centered on the cluster center at $z = 0$.
Next, we divide the tagged particles according to the density perturbation 
which is produced by additional shorter wave length components.  
The mass of a high resolution particle is 1/64 of low resolution one. 
As a result, the total number of the particles becomes $\sim 10^6$.
Our analyses are operated only for the high resolution particles.
Mass of the high resolution particle is $m \simeq 10^9 M_{\odot}$ , and its 
softening length, $\epsilon$, is set to $5$ kpc.  

\subsection{N-body calculation}
To follow the motion of the particles, we use a tree-code 
(Barnes \& Hut 1986) with the angular accuracy parameter $\theta = 0.75$, 
and we include quadrapole and octupole moments in the expansion of the 
gravitational field.  

The numerical calculation is started from redshift $z = 20$ 
and it is integrated by using the individual time step 
(Hurnquist \& Katz 1989).  
A time step for particle $i$ is given as
\begin{equation}
\triangle t_i = C \left(\frac{\epsilon^2}{a_i}\right)^{1/2},
\end{equation}
where $C$ is a constant and $a_i$ is  acceleration of 
particle $i$. This constant,  $C$, is set to 0.25.
In this case, errors in total energy is less than 1 \% through 
our simulations.  

\subsection{Halo identification}
Finding galactic halos in dense environments is a challenging work. The most 
widely used halo-finding algorithm called the friends-of-friends algorithm
(e.g., Davis et al. 1985) and the spherical overdensity algorithm 
(Cole \& Lacey 1994, Navarro, Frenk, \& White 1996) are not 
acceptable (Bertschinger \& Gelb 1991), 
because they cannot separate substructures within large halos. 

DENMAX algorithm (Bertschinger \& Gelb 1991; Gelb \& Bertschinger 1994) makes 
significant progress, but requires a substantial amount of CPU-time in 
actual calculations. Since we search halos a lot of  times through our 
simulations, we adopt lighter numerical procedure with good performance.  
Therefore, we use the adaptive friends-of-friends algorithm (Suto et al. 1992; 
Suginohara \& Suto 1992; van Kampen 1995) which enables us to avoid the 
problem in the friends-of-friends by using local densities to determine 
local linking lengths. Moreover, we remove unbound particles from halos found 
by this algorithm. This procedure is important for galactic halos in groups 
and clusters. 

In our adaptive friends-of-friends algorithm, a local linking length, 
$b_{ij}$, is calculated as follows,
\begin{equation}
b_{ij} = \beta \times \min \left[L_p, 
\frac{\rho_i(r_s)^{-1/3} + \rho_j(r_s)^{-1/3}}{2} \right],
\end{equation}
where 
\begin{equation}
\rho_i(r_s) = \frac{1}{(2 \pi r_s^2)^{3/2}} \sum^N_{j=1} 
\exp \left(-\frac{|\mbox{\boldmath $r$}_i - 
\mbox{\boldmath $r$}_j|^2} {2 r_s^2}  \right),
\end{equation}
$L_p$ is the mean particle separation,
 $r_s$ is the filtering length to 
obtain a smoothed density field and $\mbox{\boldmath $r$}_i$ is 
the position of the particle $i$. 
We specify a combination of the value of two parameters, $\beta$ and $r_s$, 
in our algorithm as follows. 
For $\beta$, we require that our algorithm is equivalent to 
the conventional friends-of-friends in the field region, so that $\beta$ 
is set to 0.2 which corresponds to the mean separation of particles in 
a virialized object.
The filtering length, $r_s$, should be determined depending on the size of 
objects in which we are interested. 
Thus, it must be larger than the size of galactic halos and smaller than the 
size of clusters. In this paper, we set it to $1$ Mpc after several tests. 

After identifying galactic halos, we remove the unbound particles.
At first, we compute the potential, $\phi_i$, for each particle $i$ 
due to all members of the halo:
\begin{equation}
\phi_i = \sum^{N_{\rm h}}_{j \ne i}\ \phi(r_{ij}),
\end{equation}
where $N_h$ is the number of particles belong to the halo.
We then iteratively remove unbound particles as follows. We compute the 
energy $E_i = (1/2) m \ |\mbox{\boldmath $v$}_i - \mbox{\boldmath $v$}_{\rm h}
|^2 + \phi_i$ for each particle in the halo, 
where $\mbox{\boldmath $v$}_{\rm h}$ is the mean velocity of the member 
particles. 
We then remove all particles with $E_i > 0$. 
The procedure is repeated until no more particles are removed.
Finally, it is identified as a galactic halo when it  contains more particles 
than the threshold number, $n_{\rm th}$, which is usually set to 15 in this 
paper. We show some tests on our halo-finding algorithm in Section 3.1.

\subsection{Creation of merging history trees of galaxies}
Our method to create galaxy merging history trees resembles to the method 
which was used by Summers et al. (1995). The main improvement is that we 
trace ``halo stripped galaxies" as well as galactic halos because halo 
disruption is probably due to insufficient resolution 
(Moore, Katz, \& Lake 1996) and lack of dissipative processes 
(Summers et al. 1995). 

To follow the evolution of galaxies in our simulation, we identify the 
galactic halos at 26 time stages with a 0.5 Gyr time interval. 
The most bound three particles in each galactic halo are tagged as tracers.
We consider three cases to follow their merging histories. 
First, for a galactic halo at a time stage, $t_{i+1}$,
where $i$ is a number of time stage, if the halo has 
more than two tracers which were contained in the same halo at the previous 
time stage, $t_i$, then the halo at $t_{i+1}$ is a ``next halo" of the halo 
at $t_i$. In this case, the halo at $t_i$ is an "ancestor" of the halo at 
$t_{i+1}$. 
Next, we consider the case that some halos at $t_{i+1}$ have one of three 
tracers of a halo at $t_i$, the halo which has the tracer that was  more 
bound in the halo at 
$t_i$ is defined as the ``next halo" of the  halo at $t_i$. 
Finally, we consider the final case.
When none of three tracers of a halo at $t_i$ are contained in any halos at 
next time stage ($t_{i+1}$), 
we define the most bound particle in this halo at $t_i$ as a ``stripped tracer". 
Then, we call both of the halos and the stripped tracers the ``galaxies" 
throughout this paper.

In this way, we construct merging history trees of galaxies. 
In order to estimate mass of stellar component of a galaxy, 
we assume that the mass of the stellar component is proportional to 
the sum of the masses of its all ``ancestors" 
(hereafter, we call this mass the "summed-up-mass"). 
Except for the case in which a large fraction of the stellar component of the 
galaxy was stripped during the halo stripping, this assumption may be valid. 
To consider mass increase due to accretion of dark matter to the halo 
after its first identification, 
we replace the summed-up-mass with the halo mass  
when the summed-up-mass is smaller than the halo mass. 

The reason using three tracers for each halo is to avoid possibility that 
we select an irregular tracer which happens to appear near the density peak 
of the halo. 
However, for almost all halos we get the same result even if we use a single 
tracer for each halo. Therefore, three tracers are enough to avoid this 
possibility. 

\section{Results}
\subsection{Galactic halos in N-body simulation}
In this subsection we show the results of some tests of our halo-finding 
algorithm to show its features and  to check its reliability.  

First, we present the distribution of dark matter and halos in the simulated 
cluster at $z = 0$ in Fig. \ref{halos}. 
The upper panel is a $x-y$ projection of a density map in a cube with sides 
$2 \times r_{200}$ ($r_{200}$ is the radius of the sphere having overdensity 
$\delta = 200$) centered on the cluster. 
Gray scale represents logarithmic scaled density given by the SPH like method 
with neighbouring 64 particles (Hernquist \& Katz 1989).  
The $x$-$y$ projection of the particles contained in galactic halos 
identified by our halo-finding algorithm is plotted in the lower panel
in Fig. \ref{halos}. 
It is found from Fig.1 that many galaxy size density peaks  survive even 
in the central part of the rich cluster and our halo-finding algorithm 
can pick up these peaks as halos. 

Next, we compare the density profiles of the halos in the simulation A 
(hereafter we refer them to field halos) with the density profile proposed by 
Navarro, Frenk, \& White (1996) (hereafter NFW).
The NFW profile approximates profiles of virialized objects obtained by 
cosmological N-body simulations well and it is written as follows:
\begin{equation}
\rho(r) = \frac{\rho_{\rm c} \delta_{\rm c}}
{\left(\frac{r}{r_{\rm s}}\right)\left(\frac{r}{r_{\rm s}} + 1\right)^2},
\label{nfw}
\end{equation}
where $\rho_{\rm c}$ is the critical density of the universe,
\begin{equation}
\delta_{\rm c} = \frac{200}{3}\frac{c^3}{\ln(1+c)-c/(1+c)},
\end{equation}
and
\begin{equation}
r_{\rm s} = r_{200}/c.
\end{equation}
In Fig. \ref{field}, we plot the density profiles of the field halos 
obtained by our halo-finding algorithm (plus signs), the profiles 
based on the spherical overdensity algorithm for $\delta = 200$ (crosses), and 
the NFW fits for latter profiles (solid lines). 
We find that the halos identified by our halo finding algorithm are well 
fitted by the NFW model except for very massive ones, and these massive halos 
have smaller radii than $r_{200}$ because it separates the dominant halos and 
their companions. 
It is also found that these halos have cores and their sizes 
are comparable to the softening length, $\epsilon$. 
potential. These cores are numerical artifacts due to the softened potential 
and they make it easier to disrupt these halos by tidal force. 
%
%

As we mentioned above, our method can pick up galaxy size density peaks 
even in the cluster environment and most selected halos without substructures 
in the field show the NFW profiles. 
However, since this method is improved only to avoid 
the clouds-in-clouds problem, we should use an alternative independent 
method for cluster halos when we argue their radii and outer density profiles. 
We can define a radius of a halo within the cluster 
using the halo density profile $\rho(r)$, where $r$ is the distance 
from center of the halo, and measuring the radius at which 
$\rho(r)$ flattens due to dominance of the cluster back ground density 
(Klypin et al. 1997; Ghigna et al. 1998). 
At the radius where the density profile is flattened, the circular velocity, 
$v_{\rm c} = (GM(r)/r)^{1/2}$, profile turns around and increases 
(Ghigna et al. 1998). 
The radius  where $\rho(r)$  flattens and one where $v_{\rm c}$ takes a 
minimum value are essentially equal (see Fig. \ref{gpv}).  
Therefor, we refer the radius at which $v_{\rm c}$ takes a minimum value 
as a radius of a cluster halo.  
It should be noted that this method allows overlap of halos, 
that is, if we estimate the mass of a halo by this method, 
mass of a halo includes the mass of the satellite's halos. 
Therefore, we cannot determine the mass of a halo by the $v_{\rm c}$ method.   

Our halo finding algorithm seems to underestimate the extent of the cluster 
halo comparing with that obtained by the $v_{\rm c}$ method.  
Does this feature cause serious problems in estimating summed-up-mass of 
galaxies?
Before cluster size objects form, we can estimate their size correctly, 
because such environment is similar to the field environment and our finding 
algorithm gives reasonable halos in the field (Fig. \ref{field}). 
After they fall into the cluster, there are three way to increase their 
summed-up-mass, that is, merging with other halos, merging with 
stripped tracers, and accretion of dark matter particles. 
Since our method identifies halos according to density peaks, 
we can treat merging of halos (i.e. peaks) properly independent of their 
size. Only when stripped tracers are enough near a density peak of a halo, 
we should regard this as merging, therefore underestimate of the extent of 
halos may not matter. 
When cluster halos increase their mass by accretion of dark matter, we 
cannot estimate increase of their summed-up-mass properly, however, 
such case may be rare, because the size of halos diminished by tidal 
interactions in the cluster as we will show in Section 3.4.3.  
Thus, we conclude that we can estimate the summed-up-mass of the cluster 
halos by our method. 

\subsection{Evolution of the whole cluster}
We define a sphere having mean over density, 200, as a virialized object, 
and we show mass, $M_{200}$, and radius, $r_{200}$, of 
the most massive virialized object at each time stage in the simulation B 
in Table \ref{tbcl}.
It is found that a cluster size object begins to form from redshift 
$z \simeq 1$, therefore we call this object a ``cluster" after $z \simeq 1$.
Indeed, the main clump of the cluster has already formed at $z = 1$ and 
it does not undergo major merging after $z = 1$ (Fig. \ref{snap}). 
We define the formation redshift, $z_{\rm form}$, of the final cluster 
(cluster at $z = 0$) as the redshift when it has accreted half of 
its final mass (Lacey \& Cole 1993), thus its formation epoch is 
$z_{\rm form} \sim 0.15$ (see Table \ref{tbcl}).  

The density profiles and the velocity dispersion profiles of the cluster 
are shown in Fig. \ref{cpr} and Fig. \ref{csig}, respectively. 
The distribution of the dark matter inside $r_{200}$ changes 
little with time (see thin lines in the upper panels of Fig. \ref{cpr} 
and Fig \ref{csig}).
It agrees the fact that this cluster evolves mainly by the accretion of dark 
matter and small clumps (Fig. \ref{snap}). 
The density profile at $z = 0$ (thin solid line) is well fitted by the NFW 
model (thick solid line) except for the central cusp ($r < 100$ kpc) 
where its slope ($\rho_{\rm cusp}(r) \propto r^{-1.35}$) 
is steeper than the NFW profile ($\rho_{\rm cusp}(r) \propto r^{-1}$) and 
well consistent with that obtained by Moore et al. (1998) 
who give  $\rho_{\rm cusp}(r) \propto r^{-1.4}$. 
In our case, the softening length, $\epsilon = 5$ kpc, is much smaller than 
$r_{s} = 300$ kpc and, moreover, the number of the particles which are inside 
the virial radius of the cluster is about two order of magnitude lager than 
that of the NFW's simulation. Thus, we conclude that we have enough resolution 
to argue density profile of central cusp ($20 < r < 100$ kpc). 

The number density profiles of halos in the cluster are plotted in the lower 
panel of Fig. \ref{cpr} (thin lines).
The number density of the halos decreases with time, especially in the central 
part of the cluster. 
The thick solid line and the thick dashed line denote the dark matter 
density, $\rho_{\rm d}$, and the number density of galaxies 
which consist of both halos and stripped tracers 
and these values are normalized by the values at $r_{200}$,  
respectively. We can see that the halo distribution is ``antibiased" 
with respect to the dark matter distribution. It is because that 
a softened halo has a core with $r_{\rm core} \sim \epsilon$, therefore, it is 
rapidly disrupted by the encounters with other halos and the tidal field of 
the cluster when $r_{\rm tidal} < 3-4 \times \epsilon$ 
(Moore, Lake \& Katz 1998). This scale is enough small for large halos 
($M_{\rm h} > 10^{11} M_{\odot}$), thus when such halos are disrupted, 
we can say that they are stripped significantly.  
Since such disruption is an artificial numerical effect and due to lack
of physics (i.e, lack of dissipational effects), 
we expect that if we perform simulations with infinite resolution or with  
baryonic component, the number density of galaxies is similar to that of
galaxies obtained here, which we get by assuming that no galaxy is disrupted 
completely. 
The number density of the galaxies at $z = 0$ (thick dashed) 
shows no ``bias" respect to the dark matter density except 
for the central part of the cluster where a central massive 
halo dominates (Fig. \ref{halos}). 
This result differs from van Kampen (1995) who suggested  that 
galaxies are more concentrated than dark matter. 
We guess that their result was the artifact produced by the artificial 
cooling adopted in his model. 

To show the effect of the dynamic friction and the domination of the 
central very massive halo, we plot the mass weighted 
velocity dispersion of galactic halos in the lower panel of Fig. \ref{csig}. 
In the central part of the cluster, it has smaller value than  
that of dark matter (upper panel). 
The difference of these two velocity dispersions implies that the large halos 
are slowed down by the dynamical friction and a central massive halo becomes 
dominant in this region.  
Except for the central region, the velocity dispersion of the cluster halos 
is almost same with that of the dark matter. Therefore, we cannot find 
the "velocity bias" which Carlberg (1994) has found for the simulated cluster 
galaxies. 

The dark matter velocity dispersion profile also decreases from $r \simeq 200$ 
kpc toward the center. 
This is consistent with the fact that density profile within this radius 
is shallower than the isothermal profile, $\rho(r) \propto r^{2}$. 
We interpret that the cold component in the central cusp of the cluster 
($r < 200$ kpc) is due to the contribution of the low velocity 
dispersion dark matter component which is  confined in the  potential well of 
the central dominant halo which is always placed at the center of the 
cluster. We will show some features of this halo in the next section. 

\subsection{Evolution of the central dominant halo}
In the simulation B, the most massive halo is always seen at the center of the 
cluster, thus we call this halo the "central dominant halo" (CDH).  
There is no doubt about the existence of the CDH in our simulated cluster, 
because 75 \% of the particles which were identified as the member of the 
CDH at $z \simeq 0.5$ also remains in the CDH at $z = 0$.
Remaining 25 \% of them are probably belong to the cluster. 

The mass evolution of the CDH which is identified by our halo-finding 
algorithm is presented in Fig \ref{frg}.
The mass of the CDH increases quickly from  $z \simeq 0.4$.
It always absorbs 15-30 galaxies of the former time stage.
Therefore, we can say that the CDH has evolved through merging 
and accretion.  
Arag\'{o}n-Salamanca et al. (1998) estimated that the stellar mass component 
in the brightest cluster galaxies (BCGs) have grown by a factor 4-5 for 
critical density models from $z \simeq 1$ by using the observed 
magnitude-redshift relation of the BCGs and evolutionary population 
synthesis models. 
The trend of the increase of mass of the CDH seems to be consistent with their 
result and the predictions by semi-analytic models 
(Kauffman et al. 1993; Cole et al. 1994; Arag\'{o}n-Salamanca et al. 1998). 
However, since there is ambiguity in distinguishing the component of 
the CDH from that of  the cluster and it is difficult to determine the extent 
of the CDH in our dissipationless simulation, we should perform simulations 
including hydrodynamic processes to investigate the evolution of the CDH and 
the stellar component within the CDH realistically, and that is left for 
further studies. 

\subsection{Formation and evolution of the galactic halos in the cluster}
\subsubsection{Mass functions}
It is interesting to compare the mass function of galaxies in a region 
which becomes the final cluster in the simulation B 
(hereafter ``pre-cluster" region) to that of the simulation A 
(hereafter ``field" region) before larger objects (groups and clusters) have 
formed. In the field region, since effects of tidal stripping  are negligible, 
stripped tracers are rare objects and  the summed-up-mass function of galaxies 
and the mass function of halos are almost same. 
Fig. \ref{mf21} shows that the summed-up-mass functions in both region 
at $z = 2$ are very similar except for the existence of very massive galaxies 
($m_{\rm sum} \gtrsim 10^{12} M_{\odot}$) in the pre-cluster region.  
The absence of high mass galaxies in the field region may be a 
consequence of the small volume of the simulation A. 
However, it is also likely that this difference is naturally explained by the 
peak formalism (Bardeen et al. 1986) which predicts that rare peaks at a mass 
scale that we selected as the massive halos  should be highly correlated in 
space, that is, they are likely to form in high density region at larger mass 
scale. 

It is also interesting to compare the above mass functions to the mass function 
expected from the Press-Schechter (PS) formalism (Press \& Schechter 1974, 
Lacey \& Cole 1993) and the conditional mass function (Lacey \& Cole 1993).
By the PS formula (here in the notation of Lacey \& Cole), 
the number density of halos with mass between $M$ and $M + dM$ at $z$ is:
\begin{equation}
\frac{dn}{dM}(M,t)\ dM = \frac{\rho_0}{M}f(S,\omega)
\left|\frac{dS}{dM}\right|dM,
\label{ps}
\end{equation}
where
\begin{equation}
f(S,\omega)\ dS = \frac{\omega}{(2\pi)^{1/2}S^{3/2}}
\exp\left[-\frac{\omega^2}{2S}\right]dS,
\end{equation}
$S = \sigma (M)^2$ is the variance of the linear density field of mass scale 
$M$, and $\omega = \delta_{\rm th} (1 + z)$ is the linearly extrapolated 
threshold on the density contrast required for structure formation.

The conditional mass function, that is, the number of halos with mass between 
$M_1$ and $M_1 + dM_1$ at $z_1$ that are in a halo with mass $M_0$ at $z_0$ 
($M_1 < M_0, z_0 < z_1$) is: 
\begin{equation}
\frac{dN}{dM_1}(M_1, z_1 | M_0, z_0)\ dM_1
= \frac{M_0}{M_1}f(S_1, \omega_1 | S_0, \omega_0) 
\left|\frac{dS}{dM}\right| dM_1,
\label{cond}
\end{equation}
where
\begin{equation}
f(S_1, \omega_1 | S_0, \omega_0)\ dS_1 
= \frac{\omega_1 - \omega_0}{(2\pi)^{1/2}(S_1 - S_0)^{3/2}}
\exp\left[-\frac{(\omega_1 - \omega_0)^2}{2(S_1 - S_0)}\right]dS_1,
\end{equation}
In Fig. \ref{mf0} we plot equation (\ref{ps}) with $\delta_{\rm th} = 1.69$ 
assuming the spherical collapse for the density contrast (Lacey \& Cole 1993) 
and equation (\ref{cond}) with $z_1 = 2$, $z_0 = 0$, and 
$M_0 = M_{200}(z = 0)$.
In this mass range, there is not so much difference between the PS mass 
function and the conditional mass function, and the summed-up-mass function in 
both regions show good agreement with the PS mass function at $z = 2$. 
The reason why the summed-up-mass function in the pre-cluster region 
agrees with the PS mass function better than the conditional mass function 
in the high mass range may be that our halo finding algorithm divides a large 
halo into small halos according to density peaks, thus, 
if we use the friends-of-friends or the spherical overdensity algorithm, 
this mass function may be more similar to the conditional mass function. 

To investigate effects of the cluster formation on the cluster galaxies, 
we plot the summed-up-mass function in the cluster at $z = 0$ 
in Fig. \ref{mf0}. 
Although the summed-up-mass function of field galaxies evolves similar to 
the PS theory, that of the cluster galaxies hardly evolves from $z = 2$ 
except for the existence of several very massive galaxies. 
This result implies that most of cluster galaxies have not increased 
their mass of the stellar component by merging and accretion from $z \simeq 2$ 
very much. 
These features seem to be consistent with the observed old population of 
cluster ellipticals, that is, the bulk of their stellar population has 
been formed at $z > 2$ and then passively evolved until present day
(Ellis et al. 1996), estimated from the surprisingly 
tight color-magnitude relation both at present (Bower et al. 1992) and at 
higher z (Ellis et al. 1997). However, inclusion of star formation processes 
and gas dynamics in our models is needed for more detailed investigation 
of the color-magnitude relation and the ages of cluster galaxies. 

\subsubsection{Merging of galaxies}
A halo that has more than two ancestors at the former time stage 
is defined as a ``merger remnant". In Fig. \ref{mrate}, we show 
the merger remnant fraction of the large galaxies 
($M_{\rm sum} \ge 10^{11} M_{\odot}$) as a function of redshift. 
In counting the number of merger remnants, we include the galaxies which 
have undergone minor mergers as well as major mergers because minor mergers 
also can lead starbursts (Hernquist 1989). 
We find that this fraction in the region dominated by the high resolution 
particles in the simulation B (hereafter cluster forming region) is 
larger than that in the field at high redshift ($z \gtrsim 3$) as expected 
from analytic work (Bardeen et al. 1986; Kauffmann 1996), that is,  
for a random Gaussian field, redshifts of collapse of galaxy scale density 
peaks are boosted by presence of  surrounding, large-scale overdensity. 
Therefore, the presence of larger objects at $z \simeq 2$ in the cluster 
formation region than in the field (see, Fig. \ref{mf21}) is well explained by 
the difference of merging efficiency between in the cluster formation 
environment and in the field. 
On the other hand, it is also found that after $z \sim 3$ this fraction 
in the cluster forming region decreases rapidly and becomes smaller than that 
in the field, and this fraction is always less than 10 \% inside the 
cluster's virial radius. 
This decline of the merger remnant fraction of cluster galaxies is due to 
high velocity dispersion of the larger objects, that is, if the relative 
velocity of a pare of galaxies is larger than inner velocity dispersion 
of the halos of these galaxies, they cannot merge (Binney \& Tremain 1987). 
Moreover, the stripping of halos by tidal fields of the groups and 
clusters also prevents merging of individual halos  
(Funato \& Makino 1992, Bode et al. 1994).   
Clearly, this decrease is the reason why the summed-up-mass of cluster 
galaxies has not evolve after $z \simeq 2$.

Because the large halos  preferentially merge, about 30 \% of the cluster 
galaxies with $M_{\rm sum} > 10^{11} M_{\odot}$ at $z = 0$ have undergone 
merging since $z \simeq 0.5$, while only 8\% of all the cluster galaxies have 
undergone it since $z \simeq 0.5$.

\subsubsection{Tidal stripping of halos}
To show  the effect of the tidal stripping on the galactic halos, we 
investigate whether  large halos ($M_{\rm h} \ge 10^{11} M_{\odot}$) 
at high redshift ($z \simeq 2$) are found as halos at lower redshift. 
Unless their descendants become stripped tracers for $n_{\rm th} = 10$, we 
call their descendants "surviving halos". 
If their descendants become stripped tracers, it means that they have lost 
large fraction  of their original halo mass and in such case dissipative effects 
should  become important, which are not included in our simulation.
In Fig. \ref{storsu}, we show the surviving fraction and the stripped fraction 
of such galaxies in the 0.5 Mpc radius bins from the cluster center.
At $z \simeq 0.5$, a large fraction of halos (60-100 \%) has survived 
(upper panel). On the other hand, at $z = 0$ (lower panel), 
more than 60 \% of the halos have been destroyed in the central part 
of the cluster, and  these fractions have clear correlation with the 
distance from the cluster center. Although we may have overestimated the 
stripping effect due to the softened potential for each particle 
(Moore, Katz, \& Lake 1996), lack of dissipative effects 
(Summers et al. 1995), and the feature of our halo-finding algorithm 
(see, Sec. 3.1),  we expect that these stripped galaxies are actually stripped 
significantly, because former two effects affect only at very small scale 
($r \lesssim 3 \epsilon$), our halo-finding algorithm can pick up very 
small density peaks within the cluster, and we treat only large halos here.  
 
Recently, Ghigna et al. (1998) have presented similar result to ours 
independently. However, they have shown only the evolution of the cluster 
halos from $z \simeq 0.5$. Clearly, the tidal stripping from halo formation 
epoch more important for the evolution of the galaxies. Our result shows that 
a number of cluster galaxies have already been strongly stripped their halos 
at $z \simeq 0.5$.  

Next we compare the radii of the halos, $r_{\rm h}$,  determined by the 
$v_{\rm c}$ method to the tidal radii of the halos estimated by the density 
of the cluster at their pericentric positions, $r_{\rm peri}$, which we 
calculate by the NFW fit of the cluster density profile at $z = 0$. 
The mean ratio of pericentric to apocentric radii, $r_{\rm peri}/r_{\rm apo}$,  
is 0.2, and 26 \% of the cluster halos are on very radial orbits, 
$r_{\rm peci}/r_{\rm apo} < 0.1$. In spite of the difference of mass of the 
clusters, these values completely agrees with those of Ghigna et al. (1998). 
The tidal radii of the halos, $r_{\rm est}$, are estimated by the following 
approximation, 
\begin{equation}
r_{\rm est} \simeq r_{\rm peri} \frac{v_{\rm max}}{V_{\rm c}},
\end{equation}
where $v_{\rm max}$ is the maximum value of circular velocity of a halo 
and $V_{\rm c}$ is the circular velocity of the cluster. 
In Fig. \ref{rt} we plot $r_{\rm est}$ against $r_{\rm h}$ 
for our outgoing halos that must have passed pericenter recently.  
We find that most  of our halos have larger radii than $r_{\rm est}$. 
Therefore, $r_{\rm est}$  seems to  give roughly the minimum radius of a 
cluster halo. It is implied that the most dominant process which leads the 
mass loss of the large cluster halos is not the high speed encounters with 
other halos but the tidal stripping due to the global tidal field of the 
cluster, because galaxies should have smaller $r_{\rm h}$ if the 
high speed encounters are important to mass loss of the cluster halos.

There is difference between our result and the result of Ghigna et al. (1998) 
who show much better agreement as $r_{\rm h} \simeq r_{\rm est}$  
except for the halos with $r_{\rm peri} < 300$ kpc which have tidal tails 
due to impulsive  collisions as they pass close to the cluster center. 
In our result, a number of our halos with $r_{\rm peri} > 300$ kpc also have 
larger $r_{\rm h}$  than $r_{\rm est}$. 
We note that halos are not stripped instantly. 
The tidal stripping time scale, $t_{\rm st}$, is roughly estimated as follows:
\begin{equation}
\frac{r}{R} \sim \left|\frac{d\Omega(R)}{dR}\right| \  r \ t_{\rm st}, 
\end{equation}
thus, 
\begin{equation}
t_{\rm st} \sim \frac{3}{2}\frac{R}{V_{\rm c}},
\end{equation}
where $r$ is a radius of a halo, $R$ is the distance from the center of the 
cluster, and $\Omega(R) = \frac{V_{\rm c}(R)}{R}$ is an angular velocity at $R$. Using this formula, 
$t_{\rm st}$ is about 1 Gyr at $R \simeq 1$ Mpc. 
Our cluster has formed very recently ($z_{\rm form} \sim 0.15$) due to 
its richness, that is, half of the cluster galaxies have accreted in the 
latest 3 Gyr. 
Therefore, we conclude that the difference between our result and theirs 
is due to the difference of the formation epoch of the clusters, and our halos 
with $r_{\rm h} > r_{\rm est}$ have not been stripped completely yet. 
 
It is interesting to compare the density profiles of the cluster halos 
and the NFW profile. To fit the density profiles of the cluster halos by 
eq.(\ref{nfw}),  we also use $r_{200}$ as a fitting parameter, because we do 
not obtain the $r_{200}$ of them from raw data.  
The top row of Fig. \ref{dprg} shows the density profiles of two halos 
(which are placed at (-1.6, 0.8) and (-0.8, -0.1) in Fig. \ref{halos}, 
respectively) with $r_{\rm h} > 2 \times r_{\rm est}$ 
and $r_{\rm peri} > 500$ kpc. 
We expect that the effect of stripping may be small for such halos. 
For both halos, the NFW model can produce good fits. 
However, most of halos which have $r_{\rm h} \simeq r_{\rm est}$  
and $r_{\rm peri} > 300$ kpc  have steeper outer density profiles 
than the NFW model, as shown in  middle and bottom rows in Fig. \ref{dprg}. 
Therefore, we can say that most of halos are stripped in some degree and have 
steeper outer profiles and some halos which have accreted recently to the 
cluster and which have not been stripped very much can retain their original 
shapes. 

According to the NFW's argument, the concentration parameter $c$ in 
eq (\ref{nfw}) should be higher for the cluster halos than that 
for the field halos, because halos within denser environments 
form at earlier epochs. 
Since increasing the numerical resolution causes steeper inner profiles 
(Moore et al. 1998), we choose the halos having similar resolution 
as those in the NFW simulation in both region and we plot the concentration 
parameters as a function of the $M_{200}$ of the halos in Fig. \ref{c}. 
It is found that the field halos have almost same values of the concentration 
parameter with the NFW's theory (solid line) and the cluster halos are more 
concentrated than the field halos.  
Two cluster halos having almost same values of $c$ with the NFW's theory 
are recently infalled halos (top row of Fig. \ref{dprg}), thus, it is expected 
that they formed in the lower density region than other cluster halos. 
We should note that there is some ambiguity in determining the concentration 
parameters for cluster halos because they have steeper outer profiles due to 
tidal stripping than the NFW model and it may lead the higher value of $c$.

\section{Discussion}
We investigate the formation and evolution of galaxy size dark halos 
in a cluster environment based on the high resolution N-body simulation.  
With our resolution (see Table \ref{data}) we  find a number of 
galaxy size density peaks (about 300 with $n_{\rm th} = 15$) within the virial 
radius of the cluster at $z = 0$. 
This result suggests that the overmerging problem can be much reduced 
by using high resolution simulation. 
However, even with our resolution, a large number of halos cannot survive 
, even if they have massive halos at higher $z$. 
This makes difficult to trace their merging histories which play important 
roles when we investigate evolution of cluster galaxies. 
To avoid this problem we trace halo-stripped galaxies as well as galactic 
halos by using the particles placed at local density peaks of the halos 
as tracers.  
This approach enables us to derive merging history trees of galaxies 
directly from dissipationless N-body simulations in various kinds of 
environments. 

We find the following results, which seems to relate to the 
evolution of the cluster galaxies, using this merging history tree: 
\begin{itemize}
	\item The galaxy distribution in the cluster do not show 
			either spatial or velocity bias except in the 
			central part of the cluster where the very massive halo 
			dominates.
	\item There is the very massive galactic halo at the center 
			of the cluster and a large fraction of dark matter particles 
			in the central part of the cluster are  confined 
			in the local potential well of this halo. 
			This halo has evolved through merging of the large halos 
			and accretion of dark matter. 
	\item At $z \simeq 2$, the halo mass functions both in the field and in 
			the cluster formation region are well fitted by the PS formula, and 
			there are massive galaxies in the cluster formation region more than in 
			the field. The summed-up-mass function of the cluster galaxies at 
			$z = 0$ has hardly changed from $z \simeq 2$. 
	\item In the cluster formation region, the number fraction of large galaxies 
		which have undergone mergers for the last 0.5 Gyr
		 is higher than that in the field  at high redshift 
		($z > 3$). After $z \simeq 3$, 
		this fraction in the cluster  formation 
		region rapidly decreases and become 
		lower than that in the field. In the cluster, merging is the rare 
		event and only a few massive halos has preferentially merged. 
	\item A large fraction of the massive halos 
		($M_{\rm h} > 10^{11} M_{\odot}$) 
		at high redshift ($z \simeq 2$) have  survived in the cluster 
		at $z \simeq 0.5$. However, after $z \simeq 0.5$ a large fraction of 
		these halos (more than 60 \% within 0.5 Mpc from the cluster center) 
		are destroyed by the tidal force of the cluster and 
		the fraction of the surviving halos  has 
		clear correlation with the distance from the cluster center. 
		It is also found that the halos which are stripped in some degree 
		have steeper outer density profiles than the NFW profile and 
		the halos which have recently accreted into the cluster 
		 have the density profiles 
		well fitted by the NFW model. 
\end{itemize}

The importance of mergers of individual galaxies to their evolution 
has been well investigated by numerical simulations (e.g., Burns 1989) and 
semi-analytic models (e.g., Kauffmann et al. 1993; Cole et al. 1994). 
Our cluster galaxies merged efficiently at high redshift ($z > 3$). 
On the other hand, the fraction of the galaxies which have undergone mergers 
recently (lower $z$) in the cluster formation region is smaller than that in 
the field. This difference of the way of merging may explain the difference 
between observed feature of field galaxies and that of cluster galaxies.   
Furthermore, merging is still important in the cluster at lower $z$, 
because it contributes to the increase of mass of the central dominant halo. 

In our results, clearly, the most important process which affects the 
evolution of galactic halos in the cluster is the tidal stripping due 
to the cluster potential. 
Since it diminishes the size of the cluster halos, these halos can hardly 
merge. Therefore, the summed-up-mass function of the cluster galaxies has not 
change so much since larger size objects (groups and clusters) formed. 
A possibility that the tidal stripping leads starbursts and 
the morphological tranceformation of galaxies and 
it causes the Butcher-Oemler effect and the density-morphology relation was 
suggested by Moore, Katz, \& Lake (1998).
Thus, inclusion of hydrodynamical processes and star formation in our 
numerical model is very interesting.   


For the next step, we will combine our merging history tree of 
galaxies derived from N-body simulations with population-synthesis models 
in order to make detailed comparison with the observational data and 
predictions of semi-analytic models. 
The results of this analysis are given in forthcoming paper.   

\acknowledgments

%
The authors wish to thank Prof. M. Fujimoto, M. Nagashima, and the referee  
for helpful discussions and comments. 
Numerical computation in this work was carried out on the HP Exemplar at the 
Yukawa Institute Computer Facility and on the SGI Origin 2000 
at the division of physics, graduate school of science, Hokkaido University.

\newpage
\figcaption[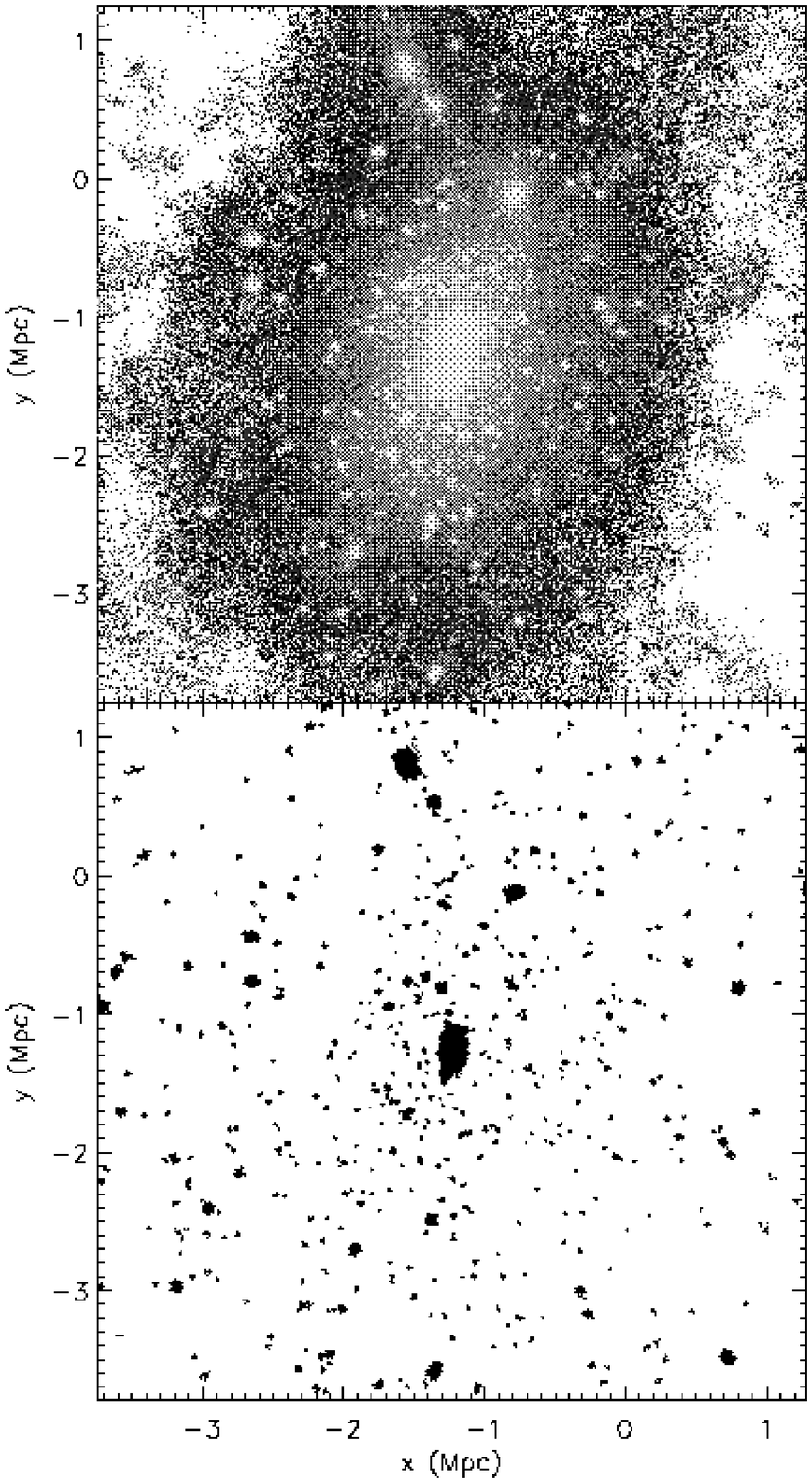]{Density map (upper panel) and the $x-y$ projection of %
the particles contained in the galactic halos (lower panel) within the %
cluster's virial radius at $z = 0$.
\label{halos}}
\figcaption[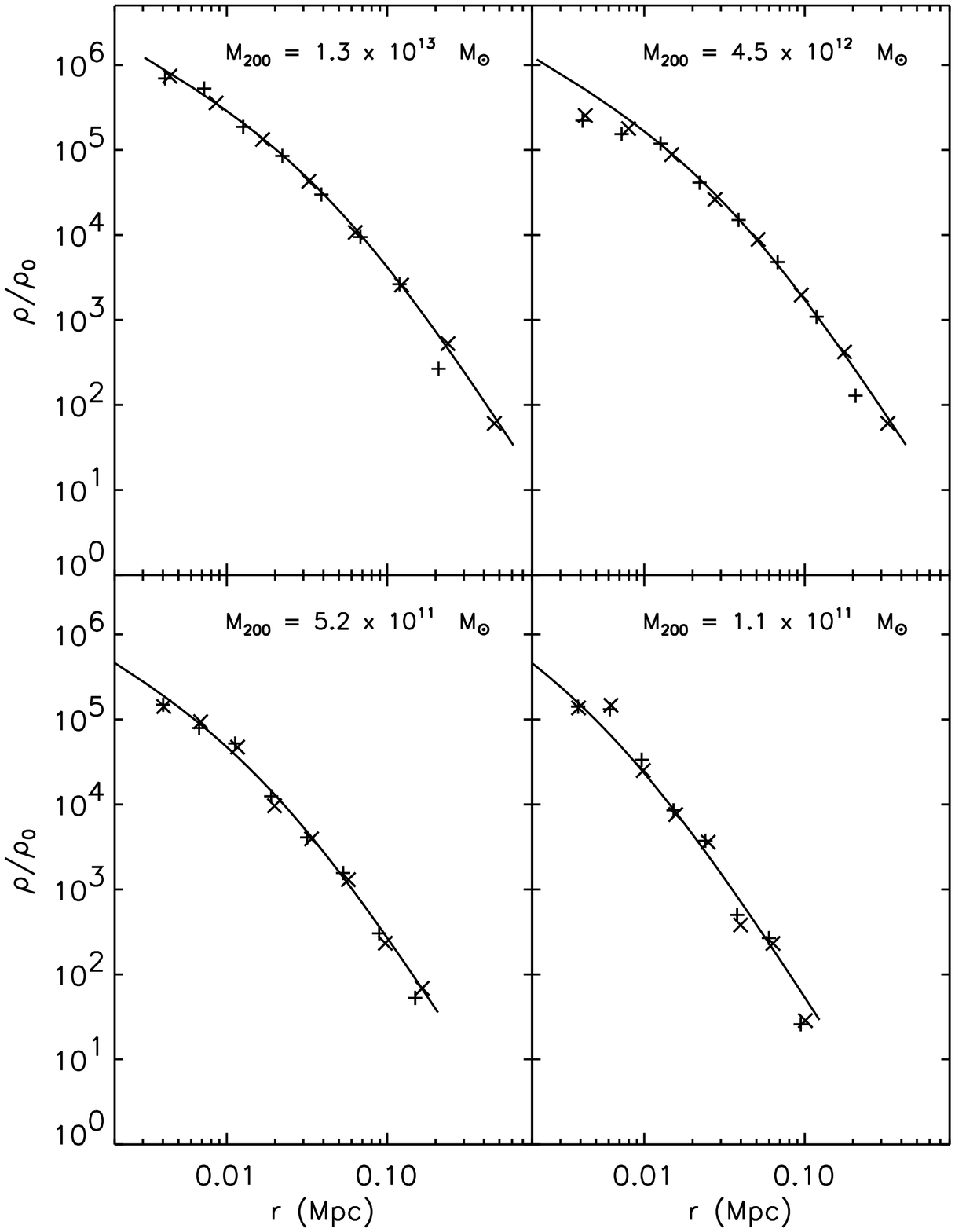]{%
Density profiles of field halos.
Pluses and crosses represent the density profiles obtained by our halo 
finding algorithm and the spherical overdensity algorithm, respectively. 
Solid lines represents the NFW fits for the profiles by the spherical 
overdensity algorithm. 
\label{field}}
\figcaption[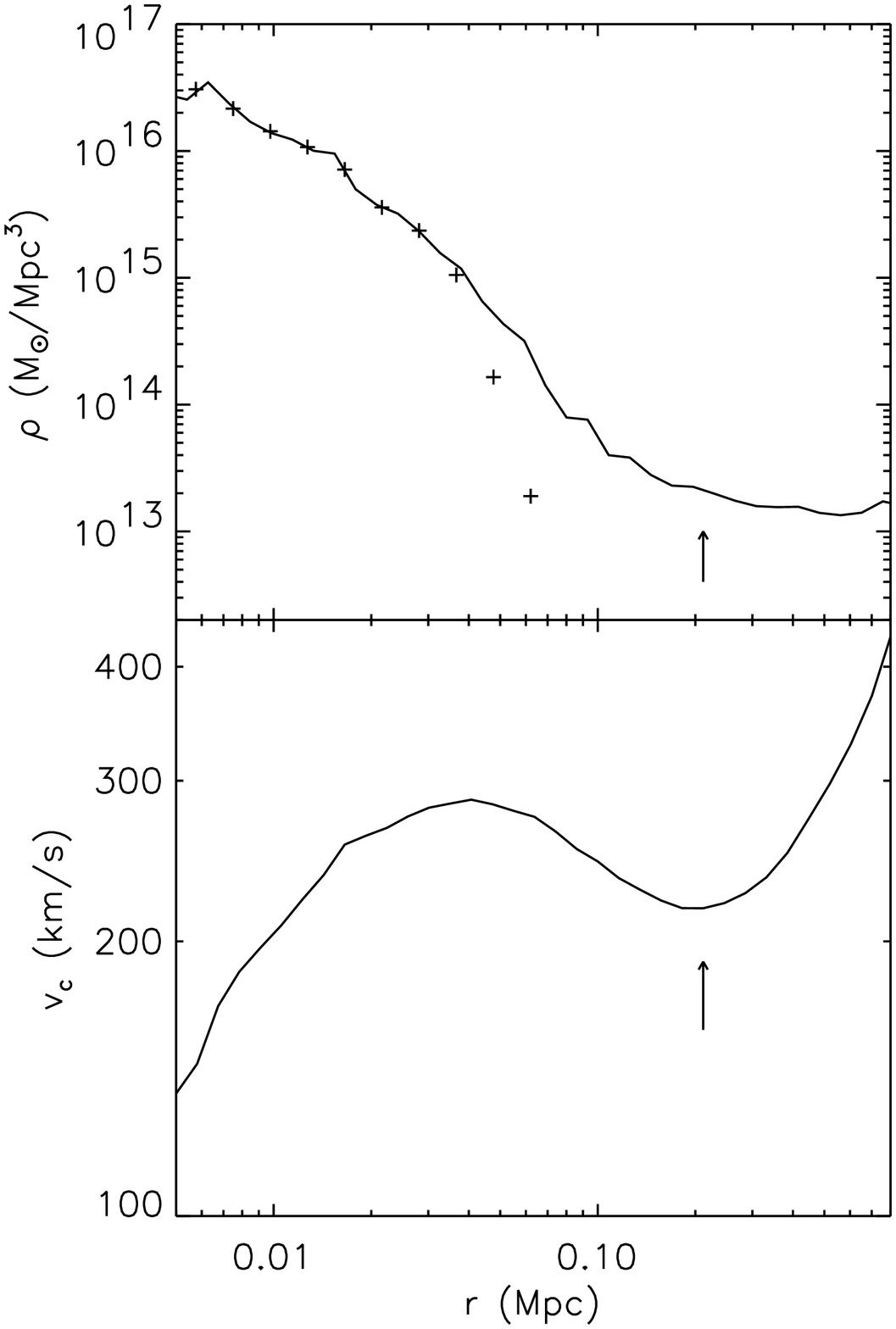]{A density profile (upper panel) and a circular velocity %
profile (lower panel) of a galactic halo in the cluster at $z = 0$. %
The density profile obtained by our halo-finding algorithm is represented by %
plus signs. The radius of the halo obtained by the $v_c$ method is indicated %
by upward arrows%
\label{gpv}}
\figcaption[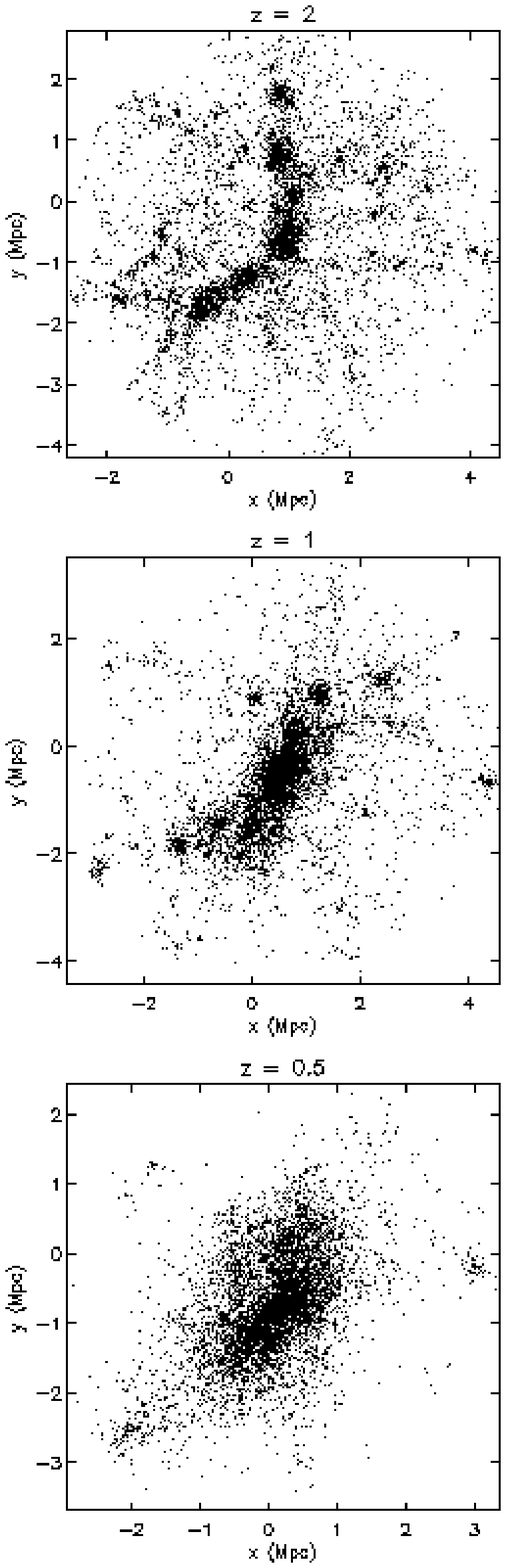]{Particle plots illustrating the time evolution of the %
cluster. One percent of the particles which are placed in the sphere having %
the same mass with the final cluster are plotted.%
\label{snap}}
\figcaption[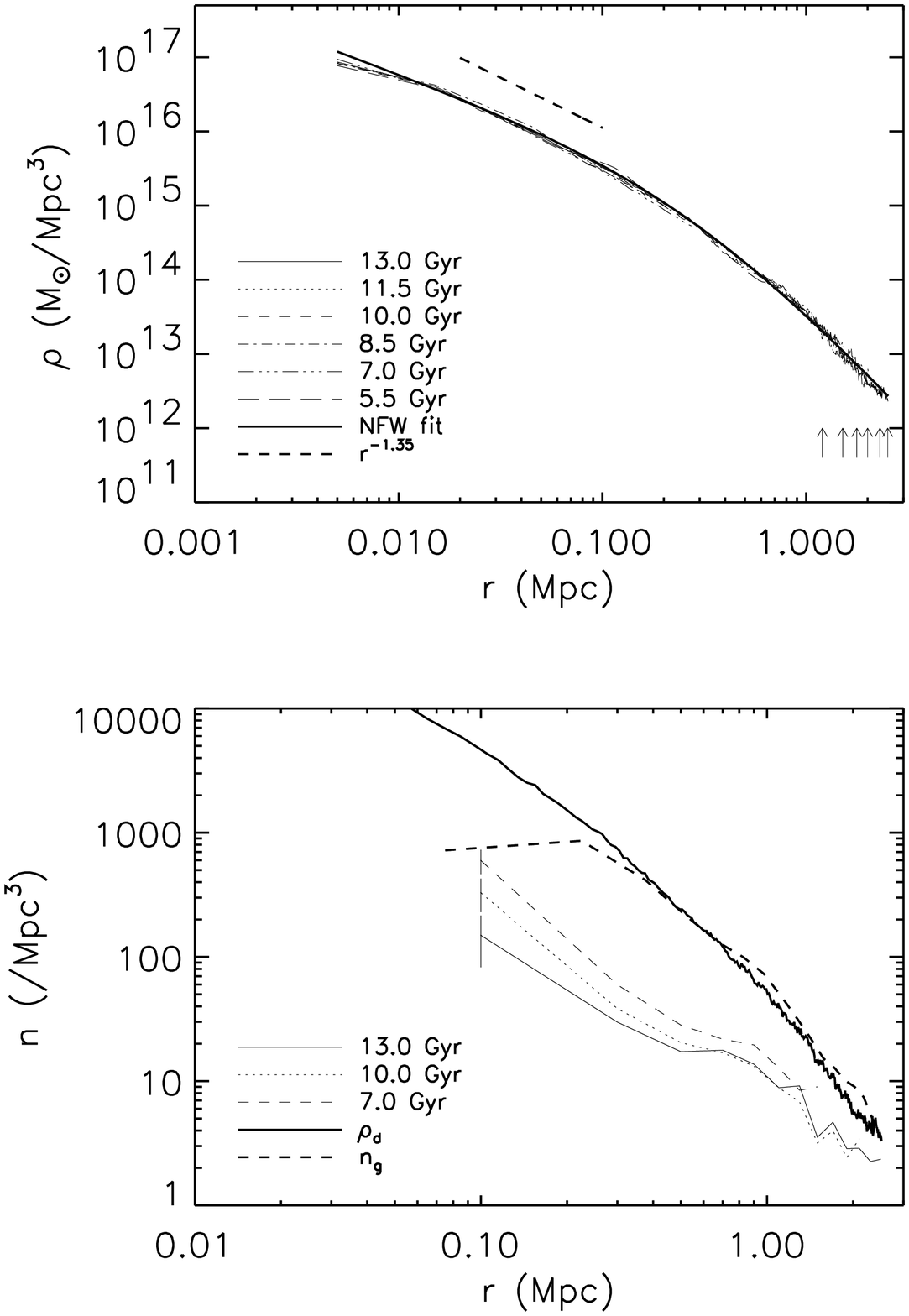]{The density profiles (upper panel) and halo number density %
profiles (lower panel) of the cluster. %
The solid line of the upper panel is the NFW fit. %
The thick solid line and the thick dashed line of the lower panel represent %
dark matter and galaxy distribution at $z = 0$, respectively. %
Both are renormalized according to their values at $r_{200}$. %
The radii of the clusters, $r_{200}$, are indicated by upward arrows.%
The errorbars are 1-$sigma$ Poissonian uncertainties estimated %
from the number of halos in the first bin. 
\label{cpr}}
\figcaption[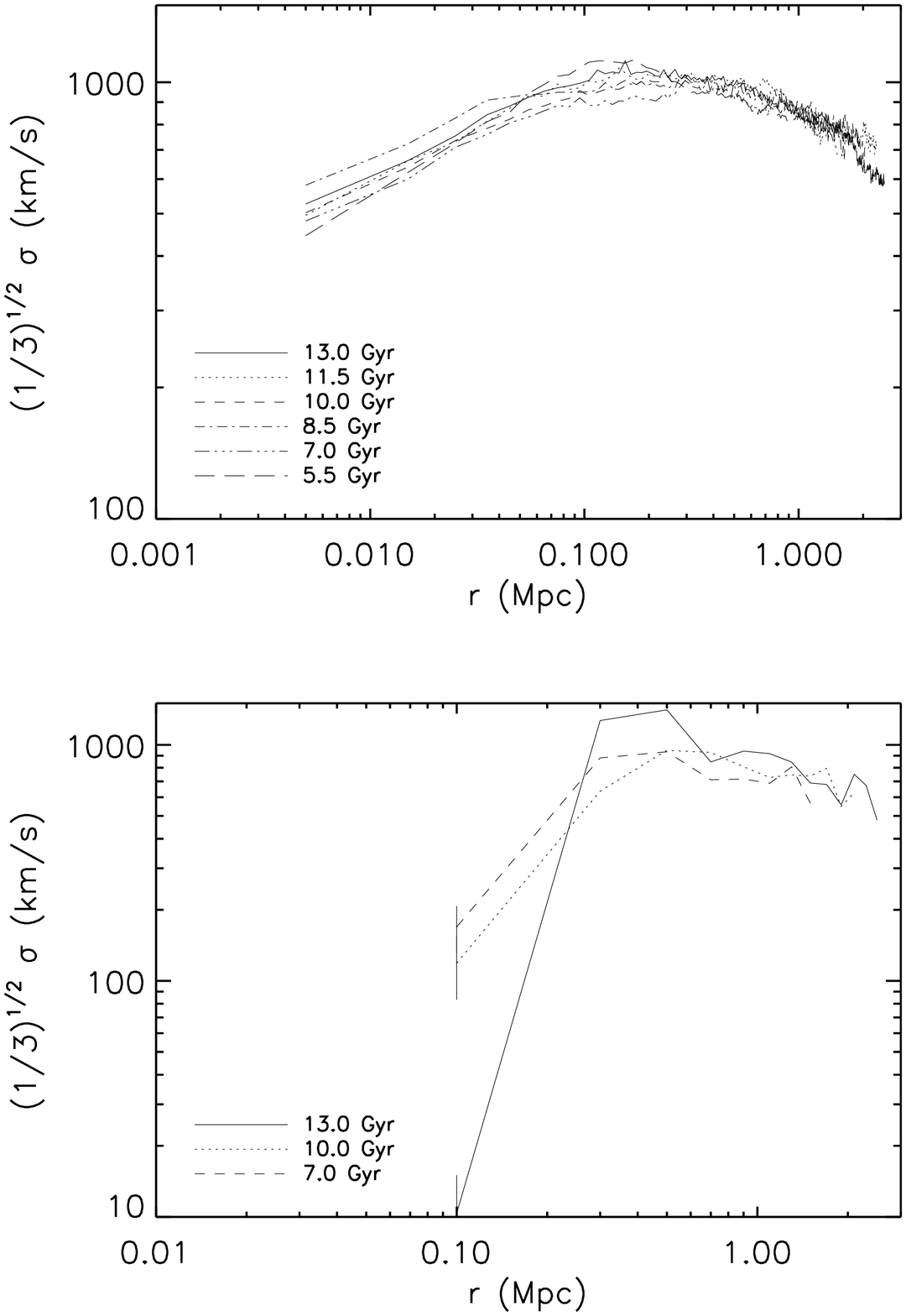]{The dark matter velocity dispersion profiles (upper panel) 
and the mass weighted halo velocity dispersion profiles (lower panel) 
of the cluster. %
The errorbars are 1-$sigma$ Poissonian uncertainties estimated 
from the number of halos in the first bin. 
\label{csig}}
\figcaption[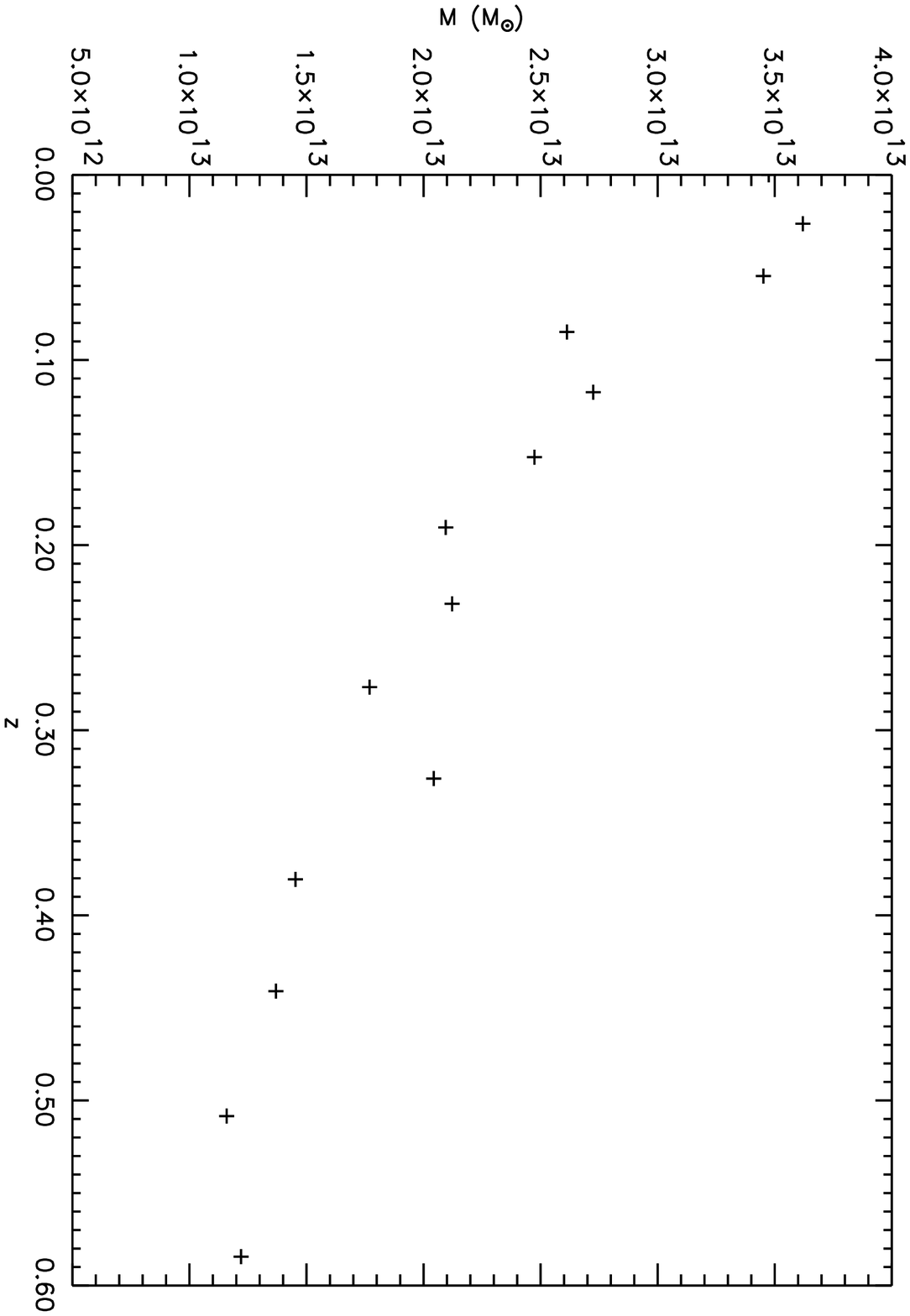]{Growth of the central dominant halo of the cluster with %
 redshift $z$.%
\label{frg}}
\figcaption[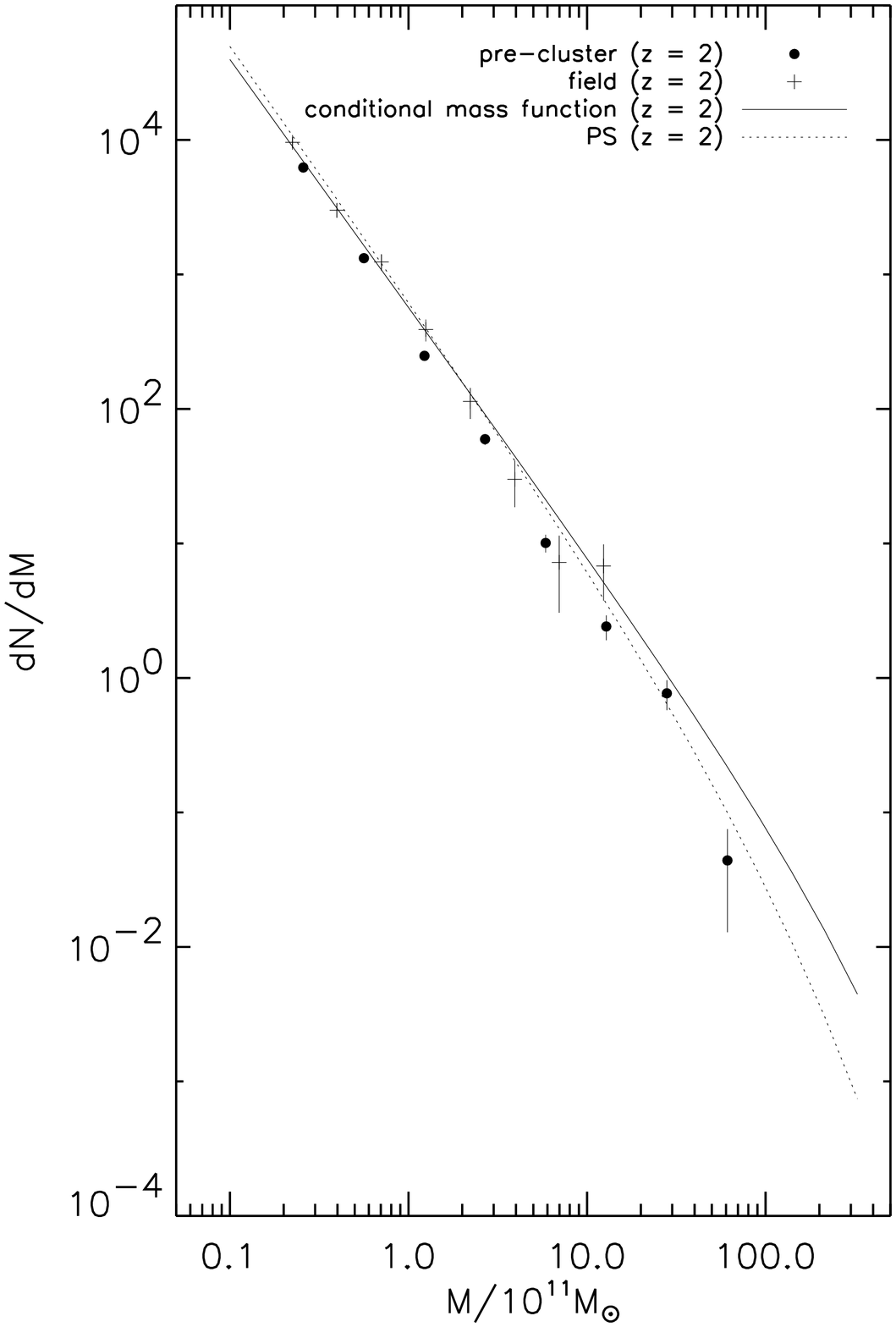]{The summed-up-mass functions in the pre-cluster region 
(filled circles) and in the field region (pluses) at $z = 2$; %
the errorbars are 1-$\sigma$ Poissonian uncertainties estimated from the 
numbers of galaxies in each mass bin. 
The solid line indicates the conditional mass function at $z = 2$ with 
$z_0 = 0$ and $M_0 = M_{200}(z = 0)$, and the dotted line indicates the PS 
mass function. The mass function in the field region and the PS mass function 
are renormalized to indicate the number of galaxies in mass $M_0$. 
\label{mf21}}
\figcaption[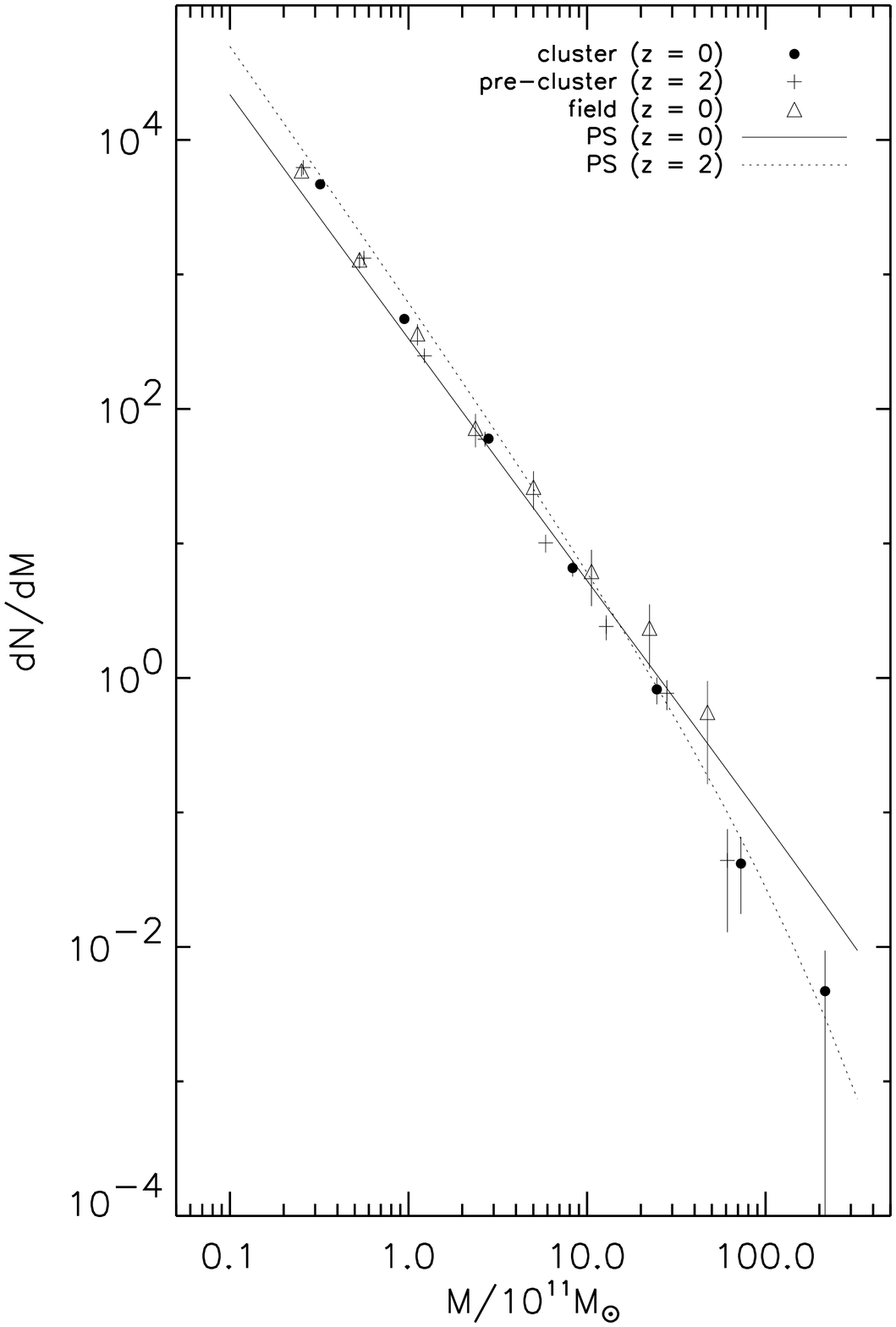]{The summed-up-mass function of galaxies in the cluster 
(filled circle) at $z = 0$. %
The summed-up-mass function in the pre-cluster region at $z = 2$ (pluses) and 
one in the field region  at $z = 0$ (triangles) are also plotted.  
The solid line and the dotted line indicate the PS mass functions at $z = 0$ 
and $z = 2$, respectively, %
\label{mf0}}
\figcaption[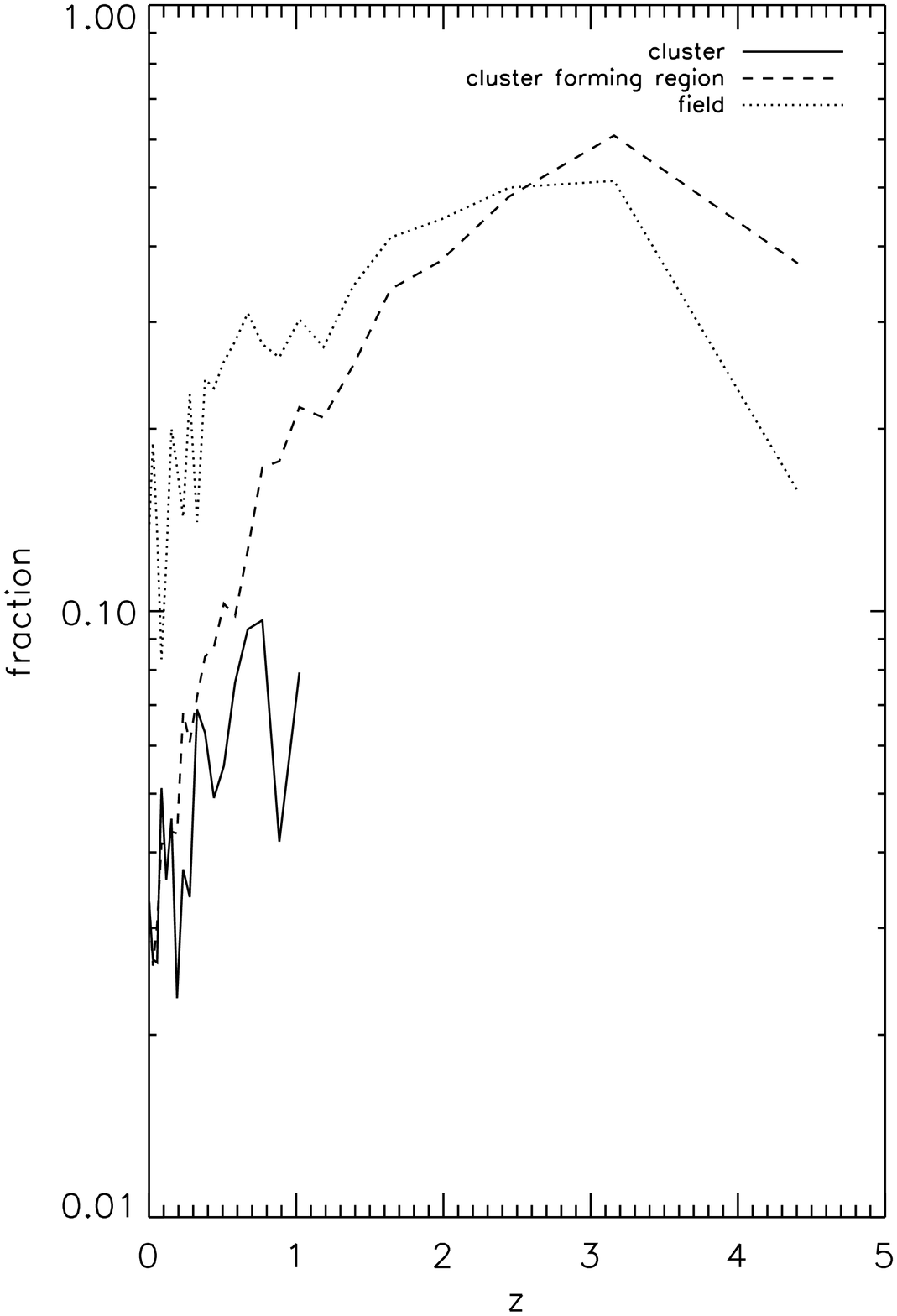]{The merger remnant fractions of the massive 
galaxies with $M_{\rm sum} > 10^{11} M_{\odot}$ in the cluster (solid line), 
in the cluster forming region (dashed line), and in the field (dotted line).  
\label{mrate}}
\figcaption[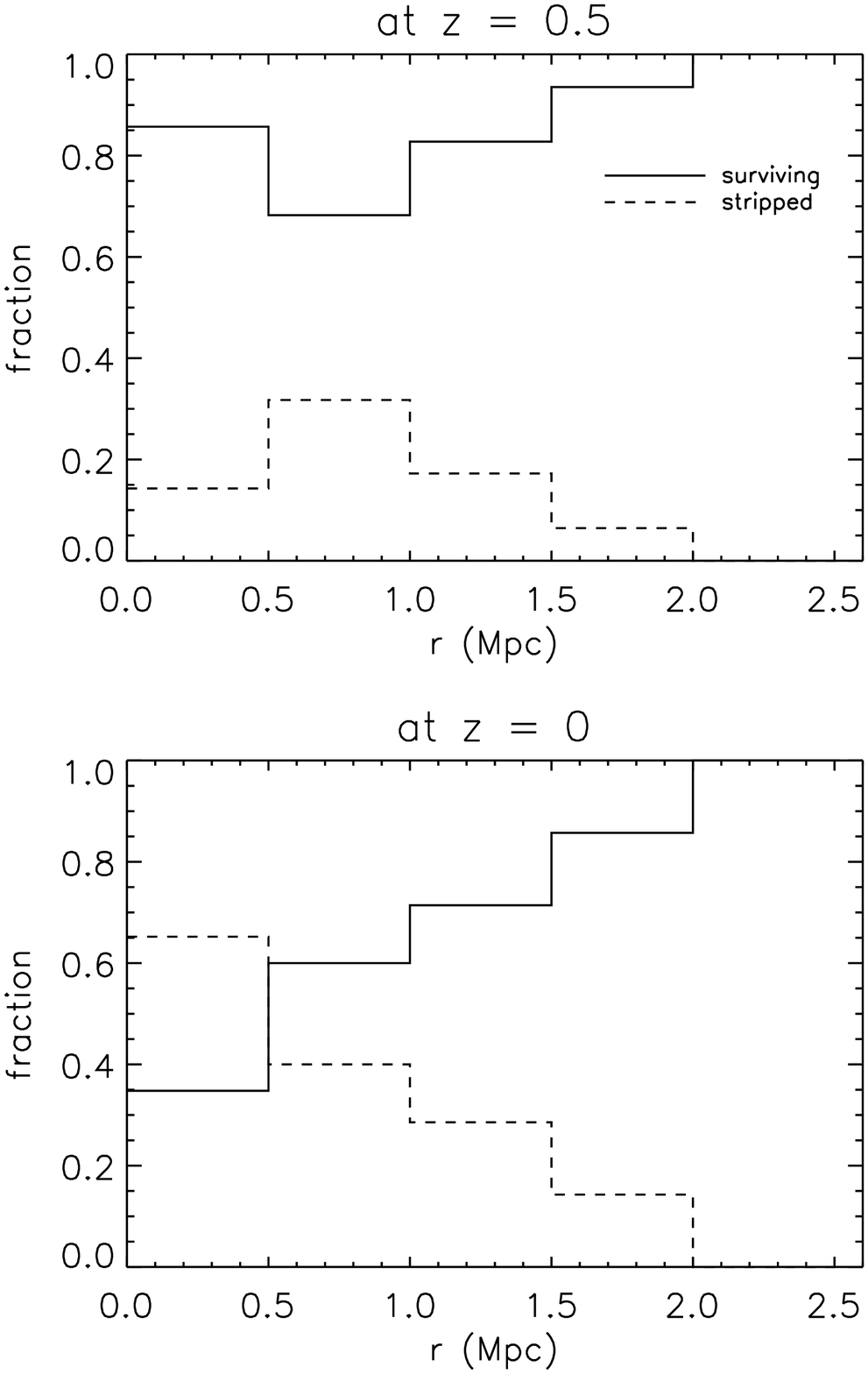]{The surviving fraction  (solid line) and the stripped 
fraction (dashed line) of galaxies which have massive halos 
($M_{\rm h} >  10^{11} M_{\odot}$) at $z \simeq 2$.  %
They are plotted in the 0.5 Mpc bins from the center of the cluster 
at $z \simeq 0.5$ (upper panel) and at $z = 0$ (lower panel). 
\label{storsu}}
\figcaption[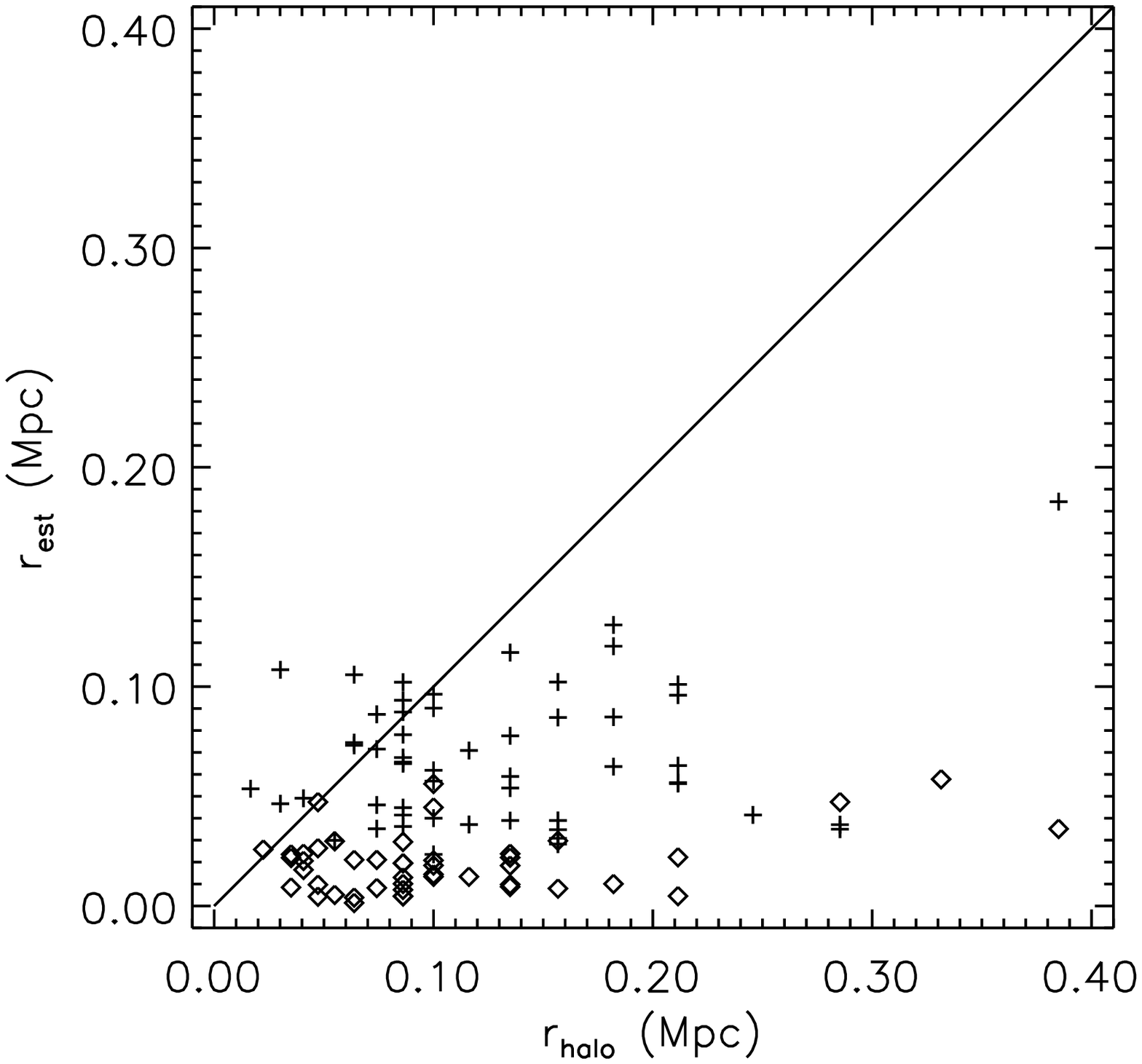]{Measured values of halo tidal radii against their %
expected values, assuming that the halos have isothermal mass distributions %
that are tidally stripped at their pericentric positions. %
The plus signs represents outgoing halos (at $z = 0$). The diamonds denote those %
with $r_{\rm peri} < 300$ kpc. %
\label{rt}} 
\figcaption[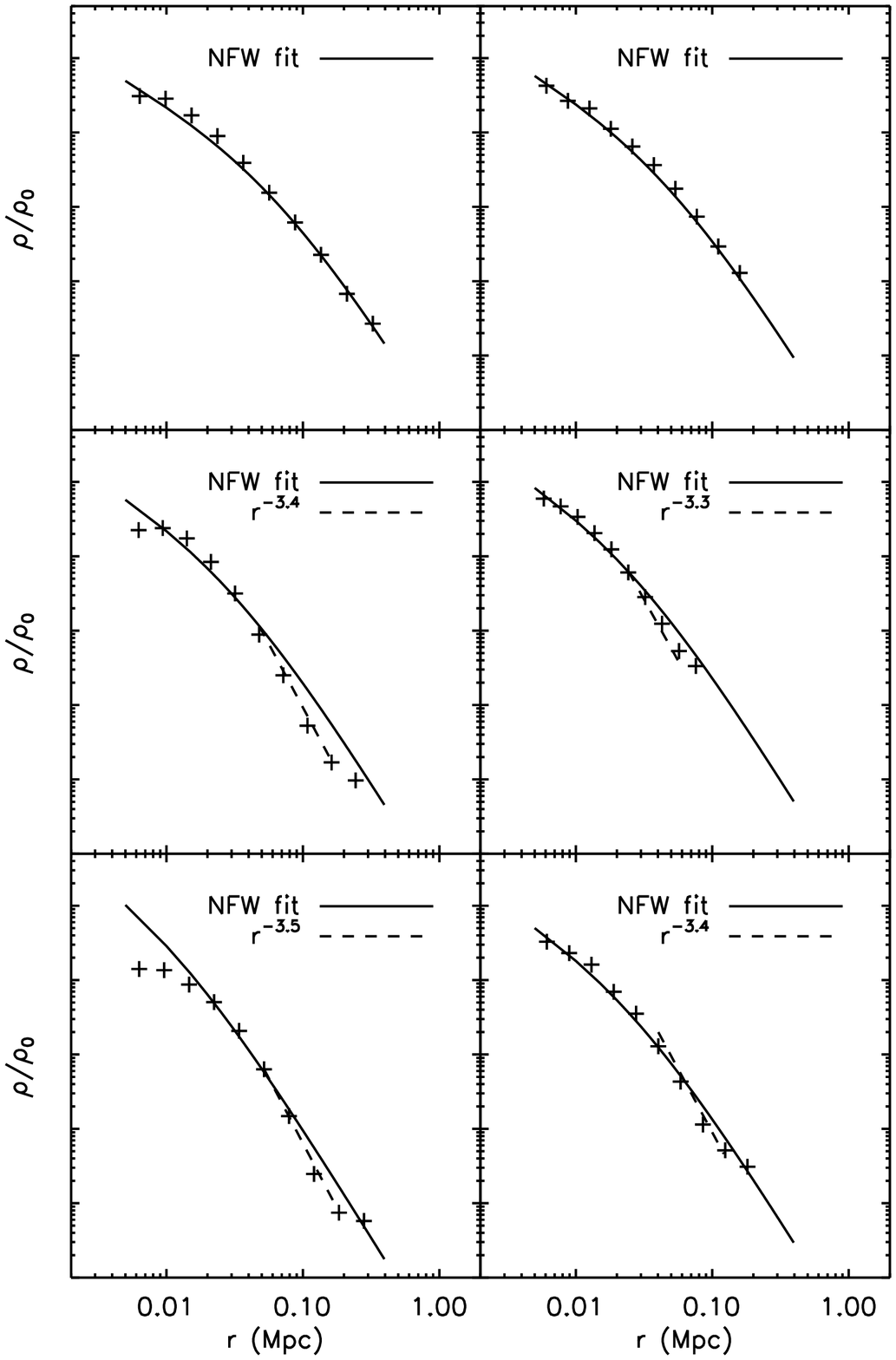]{Comparison between the density profiles (at $z = 0$) %
of  cluster halos (plus signs) and their fits by the NFW profiles 
(solid lines). %
The top row resents those of two massive halos with $r_{\rm peri} > 500$ %
kpc and $r_{\rm h} > 2 \times r_{\rm est}$. 
For other plots we show power low fits for outer profiles of the halos %
(dashed lines) with $r_{\rm h} \sim r_{\rm est}$. %
\label{dprg}}
\figcaption[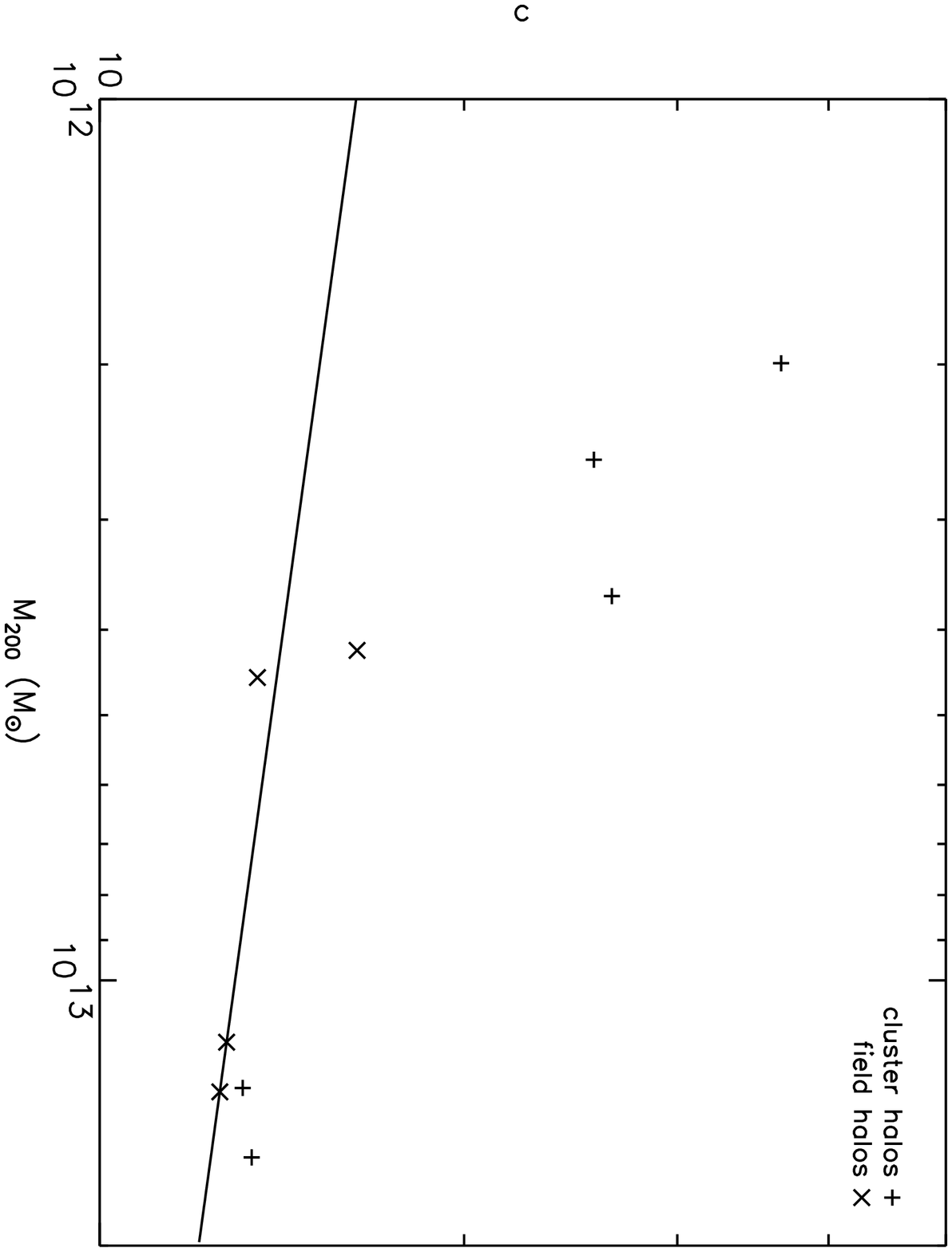]{The concentration parameters for cluster halos (pluses) 
and for field halos (crosses). The solid line denote the analytic prediction 
by NFW.
\label{c}}
\newpage
\begin{table}
\begin{center}
\begin{tabular}{cccccccccc}\hline
sim.&constraint&$N_{\rm h}$&$N_{\rm l}$&$\epsilon_{\rm h}$&$\epsilon_{\rm l}%
$&$m_{\rm h}$&$m_{\rm l}$&size&$M_{\rm most}$\\
    & & & &[kpc]&[kpc]&$[M_{\odot}]$&$[M_{\odot}]$&[Mpc]&$M_{\odot}$\\ \hline 
A&none&91911&--&5&--&$1.08 \times 10^9$&--&7&$1.6 \times 10^{13}$\\
B&$3\sigma$ peak&958592&97953&5&50&$1.08 \times 10^9$&$6.9 \times 10^{10}$&30
&$9.3 \times 10^{14}$\\ \hline
\end{tabular}
\end{center}
\caption{Parameters of two simulations. %
Subscripts h and l indicate high-resolution and low-resolution particles, %
respectively.}
\label{data}
\end{table}

\newpage
\begin{table}
\begin{center}
\begin{tabular}{rrrr}\hline
$t$ (Gyr)&redshift &$r_{200}$ (Mpc)&$M_{200}$ ($M_{\odot}$)\\ \hline
0.5&7.4&0.04&4.4 $\times 10^8$\\
1.0& 4.4& 0.10& 1.1 $\times 10^{10}$\\
1.5& 3.2& 0.17& 6.8 $\times 10^{10}$\\
2.0& 2.4& 0.30& 4.5 $\times 10^{11}$\\
2.5& 2.0& 0.39& 1.2 $\times 10^{12}$\\
3.0& 1.6& 0.51& 2.9 $\times 10^{12}$\\
3.5& 1.4& 0.60& 5.2 $\times 10^{12}$\\
4.0& 1.2& 0.72& 9.9 $\times 10^{12}$\\
4.5& 1.0& 0.87& 1.9 $\times 10^{13}$\\
5.0& 0.89& 1.02& 3.4 $\times 10^{13}$\\
5.5& 0.77& 1.21& 5.8 $\times 10^{13}$\\
6.0& 0.67& 1.32& 8.0 $\times 10^{13}$\\
6.5& 0.58& 1.42& 1.0 $\times 10^{14}$\\
7.0& 0.51& 1.51& 1.3 $\times 10^{14}$\\
7.5& 0.44& 1.58& 1.6 $\times 10^{14}$\\
8.0& 0.38& 1.68& 2.0 $\times 10^{14}$\\
8.5& 0.33& 1.77& 2.4 $\times 10^{14}$\\
9.0& 0.28& 1.85& 2.9 $\times 10^{14}$\\
9.5& 0.23& 1.94& 3.4 $\times 10^{14}$\\
10.0& 0.19& 2.03& 4.1 $\times 10^{14}$\\
10.5& 0.15& 2.13& 4.8 $\times 10^{14}$\\
11.0& 0.12& 2.21& 5.6 $\times 10^{14}$\\
11.5& 0.08& 2.31& 6.6 $\times 10^{14}$\\
12.0& 0.05& 2.38& 7.4 $\times 10^{14}$\\
12.5& 0.03& 2.45& 8.3 $\times 10^{14}$\\
13.0& 0.0& 2.52& 9.3 $\times 10^{14}$\\ \hline
\end{tabular}
\end{center}
\caption{The mass of the most massive virialized object $M_{200}$ and %
its radius $r_{200}$ at each time stage. }
\label{tbcl}
\end{table}
\end{document}